\documentclass{emulateapj}
\usepackage{graphicx,verbatim}
\usepackage{amsmath}
\usepackage{amssymb}
\usepackage{natbib}
\begin{document}
\def\etal{et al.\ \rm}
\def\ba{\begin{eqnarray}}
\def\ea{\end{eqnarray}}
\def\etal{et al.\ \rm}

\title{Ab-initio pulsar magnetosphere: three-dimensional particle-in-cell simulations of oblique pulsars.}

\author{Alexander A. Philippov\altaffilmark{1}, Anatoly
  Spitkovsky \& Benoit Cerutti}

\affil{Department of Astrophysical Sciences, 
Princeton University, Ivy Lane, Princeton, NJ 08540}
\altaffiltext{1}{sashaph@princeton.edu}


\begin{abstract}

We present ``first-principles'' relativistic particle-in-cell simulations of the oblique pulsar magnetosphere with pair formation.
The magnetosphere starts to form with 
particles extracted from the surface of the neutron star. These particles are accelerated by 
surface electric fields and emit photons capable of producing
electron-positron pairs. We inject
secondary pairs at locations of primary energetic particles, 
whose energy exceeds the threshold for pair formation. 
We find solutions that are close to the ideal force-free magnetosphere, with the Y-point and current sheet.
Solutions with obliquities $\leq 40^{\circ}$ do not show pair
production in the open field line region, because the local current density along magnetic field is below the Goldreich-Julian value. The bulk outflow in these solutions is charge separated, and pair formation happens in the current sheet and  return current layer only. Solutions with higher inclinations show pair production in the open field line region, with high multiplicity of the bulk flow and the size of pair-producing region increasing with inclination. We observe the spin-down of the star to be comparable to MHD model predictions. The magnetic dissipation in the current sheet ranges between 20\% for the aligned rotator and 3\% for the orthogonal rotator. Our results suggest that for low obliquity neutron stars with suppressed pair formation at the light cylinder, the presence of phenomena related to pair activity in the bulk of the polar region, e.g., radio emission, may crucially depend on the physics beyond our simplified model, such as the effects of curved space-time or multipolar surface fields.

\end{abstract}


\section{Introduction}
In the last 15 years the pulsar magnetosphere was intensively studied
in the magnetohydrodynamic (MHD) limit, including ideal force-free electrodynamics \citep{ckf99, gruzinov_pulsar_2005, McKinney06, tim06, spit06,kc09,petri12a}, resistive
force-free \citep{lst11, kalap12}, and full relativistic MHD \citep{kom06,SashaMHD}. While these solutions agree on the general shape of the plasma-filled magnetosphere, the origin of the magnetospheric plasma and the properties of particle acceleration necessary for production of observed radiation cannot be addressed within the MHD approach, necessitating a kinetic treatment. 

Recently, the aligned pulsar magnetosphere was studied with
first-principles relativistic particle-in-cell simulations
(\citealp{PhSp14}, hereafter, PS14; \citealp{Chen14}, hereafter, CB14; \citealp{Benoit14}). PS14 showed that
injecting abundant pairs everywhere in the magnetosphere 
drives the solution towards the force-free configuration. \citet{Benoit14} studied the transition between the disk-dome configuration
and the filled solution by varying the rate of particle
injection at the stellar surface. \citet{Beloborodov08} and \citet{Tim13} studied one-dimensional polar cap cascade and found that in the region where current density along the magnetic field, $j_{\parallel}$, is below the Goldreich-Julian value, $j_{GJ} = -\boldsymbol{\bf{\Omega_{*}}}\cdot\boldsymbol{B}/2\pi$ (here, $\boldsymbol{\Omega_{*}}$ is the stellar angular velocity vector, and $\boldsymbol{B}$ is the local magnetic field), particles in the space-charge-limited flow are not accelerated up to relativistic energies and do not produce secondary pairs. Efficient particle acceleration, and, consequently, active pair formation, should thus be confined to the regions where $j_{\parallel}/j_{GJ}<0$ or $j_{\parallel}/j_{GJ}>1$. CB14 presented the first
global aligned simulation with $e^{\pm}$ discharges and concluded that pair formation in the outer
magnetosphere is required to produce the solution similar to force-free but with strong charge separation, consistent with sub-GJ current flow in the polar cap region. This raised the question of whether high-multiplicity state of the magnetosphere is ever achieved, particularly in more realistic oblique rotators.

The goal of this Letter, therefore, is to describe the magnetospheres
of oblique pulsars. For this, we construct  
relativistic particle-in-cell (PIC) simulations of the oblique pulsar
with different prescriptions for populating the magnetosphere with
plasma, focusing on solutions with realistic prescription for pair formation. The Letter is
organized as follows. In \S
\ref{sec:numerical-models} we describe our numerical method, in \S \ref{sec:test} we demonstrate 
the test solution with abundant neutral plasma supply, and in \S \ref{sec:pulsar}
 we present our results on three-dimensional magnetosphere structure
 of aligned and oblique rotators with
pair production.

\section{Numerical method and setup}
\label{sec:numerical-models}

To simulate the magnetosphere we use the 3D electromagnetic PIC code
TRISTAN-MP \citep{anatolycode}. To a good accuracy, the star
can be described as a rotating magnetized conductor. Unipolar induction
generates electric fields corresponding to a quadrupolar surface charge. In the limit of zero work function these
charges can be pulled from the surface by the electric field, populating the
magnetosphere with plasma. Below we ignore the complications of
extracting ions from the surface and assume that extracted positive
charges are positrons. 
In addition to extraction of surface charges, we consider two scenarios for plasma supply.
First, we extend the study of PS14 to
oblique rotators by considering abundant neutral plasma
injection in the whole magnetosphere. Second, we add a more realistic prescription, which approximates pair production by magnetic conversion of photons and two-photon collisions (see \S\ref{sec:pairform}). 

We use linearity of
Maxwell's equations to represent the electric field as a
superposition: ${\bf{E}}={\bf{E_{vacuum}}}+{\bf{E_{plasma}}}$, where ${\bf{E_{vacuum}}}$ is the analytical field of
the vacuum rotator 
\citep{MicLi14},
and ${\bf{E_{plasma}}}$ is the field computed from PIC particles
\citep{Anatolychsep}. The same decomposition is done for the magnetic
field. Inside the conducting sphere,  
the plasma fields are forced to zero with a smoothing kernel, while the vacuum fields are set to corotation values.

We use a Cartesian grid of $1024^3$ cells, with stellar radius $R_{*} = 35$ cells, and  light cylinder  $R_{LC} =
120$ cells. Particles penetrating into the star by two cells are removed. The outer walls 
have radiation boundary conditions for fields and particles.

\section{Test problem: ``force-free'' solution}
\label{sec:test}
\begin{figure*}

\centering
\includegraphics[width=1.\textwidth]{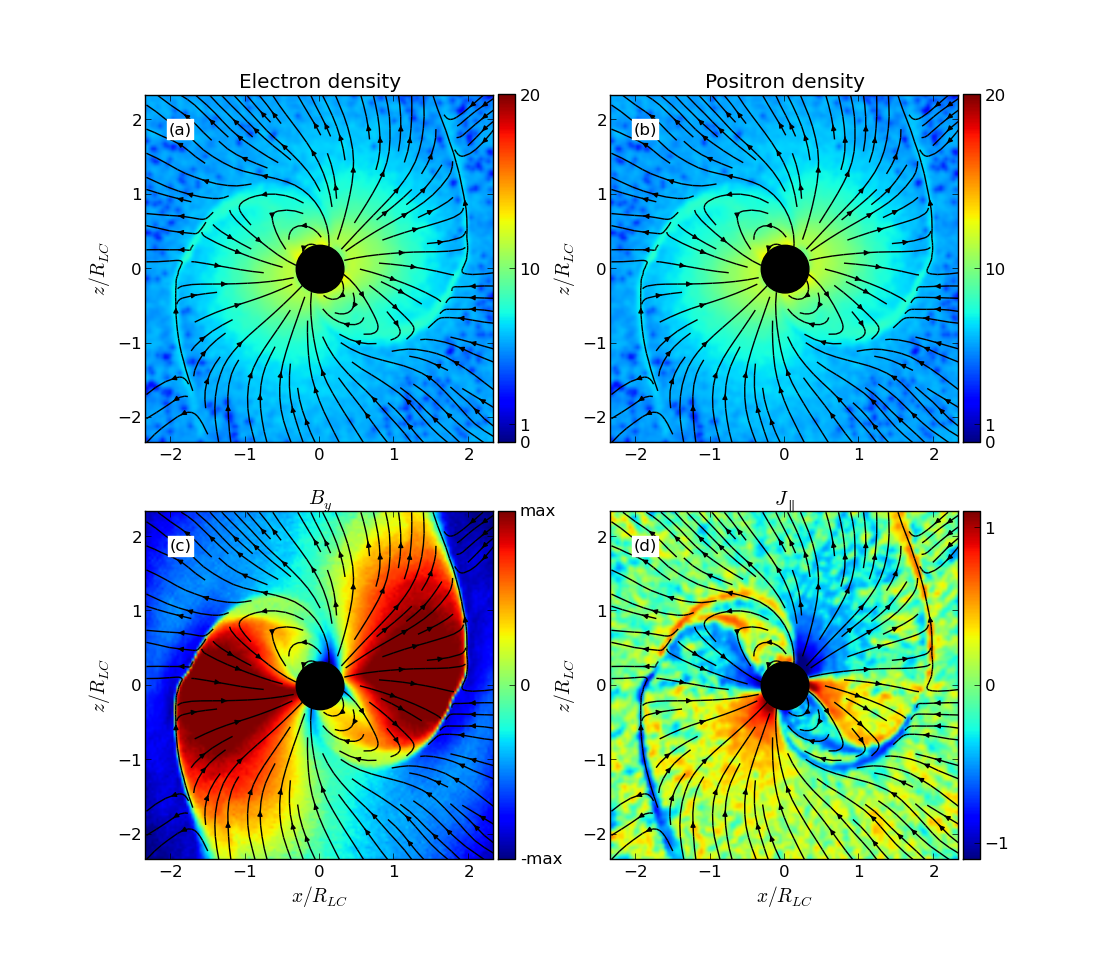}

\caption{Slice through the 
${\boldsymbol \mu}-{\bf \Omega_*}$ plane for  $\chi=60^{\circ}$ pulsar magnetosphere with abundant pair injection, shown after two rotational periods, $R_*/R_{LC}=0.3$:  (a, b)
  electron and positron density distribution, normalized by $\Omega_* B/2\pi e c$; poloidal field lines are shown in black;
  (c) out-of-plane component of the magnetic
  field (color); (d) current component parallel to the magnetic field, normalized by $\Omega_* B/2\pi$.}
\label{fig1:ffree}
\end{figure*}

\begin{figure*}
\centering
\label{fig2:aligned}
\includegraphics[width=1.\textwidth]{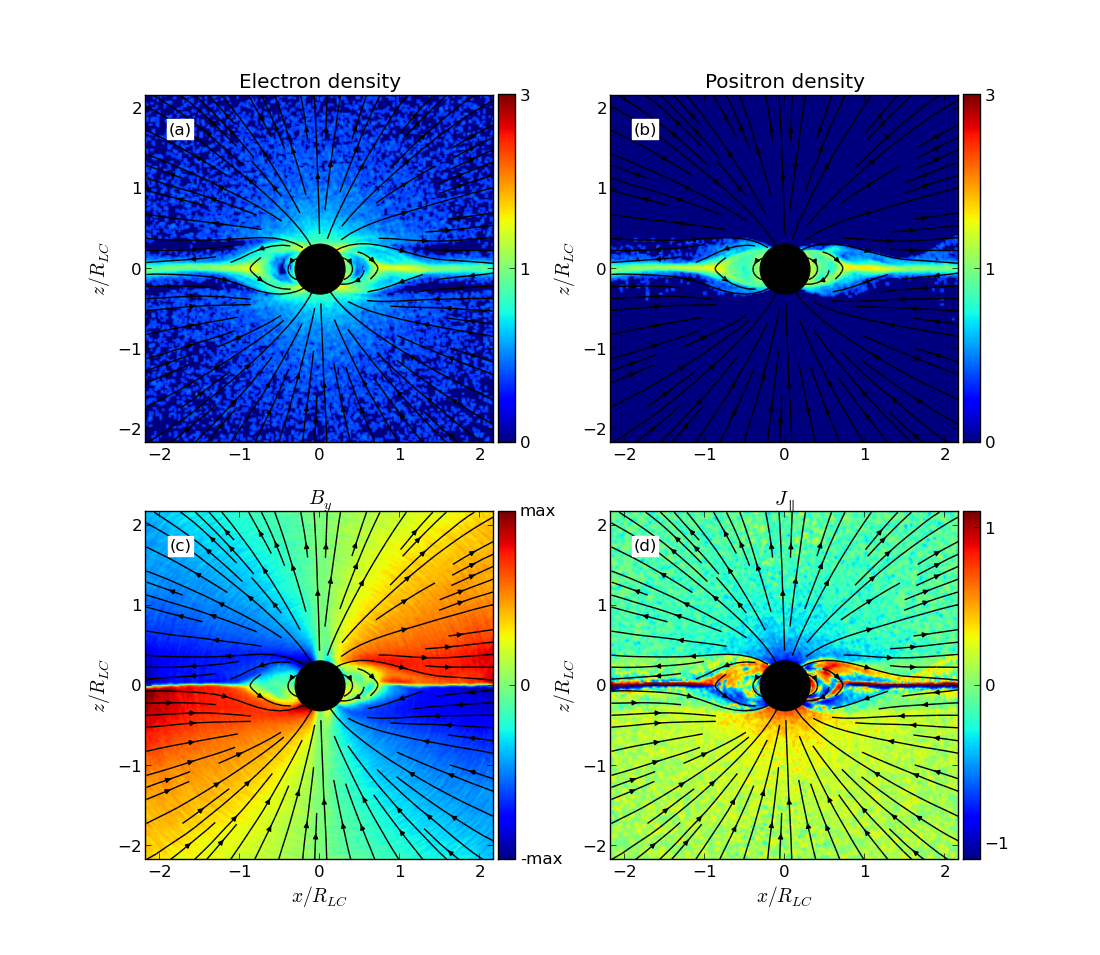}

\caption{Same as Fig.~\ref{fig1:ffree}, but for  aligned pulsar magnetosphere with  self-consistent pair formation.}
\end{figure*}
In this section we extend PS14 and build a force-free solution for the oblique rotator. Neutral pair plasma is injected with zero velocity at a fixed rate
in every cell within a sphere of $R= 2R_{LC}$, if the 
magnetization in a cell exceeds a certain limit, $\sigma =B^2/(4\pi n m_e c^2) >
1000 (R_*/r)^3$. Here, $B$ is the absolute value of the magnetic
field, and $n$ is the plasma density.
This limiter helps to prevent overloading of the magnetosphere with plasma,
particularly in the closed zone, while keeping the flow well-magnetized.
The magnetization of our solution is $\sigma\approx 1000$ near the
pole (around 20 particles per cell, local skin depth is 1.1 cells), and $\sigma\approx 20$ at the light cylinder (around 1 particle per cell, local skin depth is 5 cells).

Magnetic field in the plane defined by the magnetic moment $\boldsymbol{\mu}$ and stellar angular velocity $\bf{\Omega_{*}}$ vectors for $60^\circ$ rotator 
is shown in Fig.~1c, where color represents the out-of-plane component of the magnetic field. In
agreement with the MHD solution \citep{SashaMHD},  poloidal field lines in this plane become
purely radial beyond $R_{LC}$. The
Y-point is located approximately at the light cylinder, and the current
sheet oscillates around the equatorial plane. 
As shown in Fig.~1a,b, our kinetic
solution has high multiplicity  ($\approx 10$) in the bulk of the polar outflow, which carries sub-GJ current (see Fig.~1d).
The regions with current above the GJ value, favorable for
efficient pair formation, are the current sheet and the rims of the polar cap, where volume return current flows\footnote{The pair producing rim is located at the southern part of the polar cap (for the northern
hemisphere), unrelated to the location of ``favorably curved''
field lines \citep{arons79}.}.

The field structure is quasi-stable in the corotating frame. 
We note that within the simulation box the role of drift-kink instability of the current
sheet, pointed out for the aligned rotator by PS14 and studied by \citet{Benoit14}, decreases
with increasing obliquity, becoming negligible already for
$60^{\circ}$. This is a consequence of the decrease of conduction current in the current sheet with increasing obliquity.

We ran simulations for obliquities $30^{\circ}$, $60^{\circ}$ and $90^{\circ}$, measuring the Poynting
flux integrated over a sphere with $r = R_{LC}$. Combining with the
aligned case (PS14), we conclude that the 
spin-down energy loss of the PIC solution is consistent with
MHD studies:
\begin{equation}
L=\frac{\mu^2\Omega_*^4}{c^3}(k_0+k_1\sin^2\chi),
\end{equation}
with coefficients $k_0 = 1.0 \pm 0.1$,  $k_1 = 1.1 \pm 0.1$ (see Figure 5), and $\chi$ the inclination angle. 

\section{Pulsar magnetosphere}
\label{sec:pulsar}

\subsection{Pair creation process}
\label{sec:pairform}
\begin{figure*}

\centering
\label{fig3:60}
\includegraphics[width=1.\textwidth]{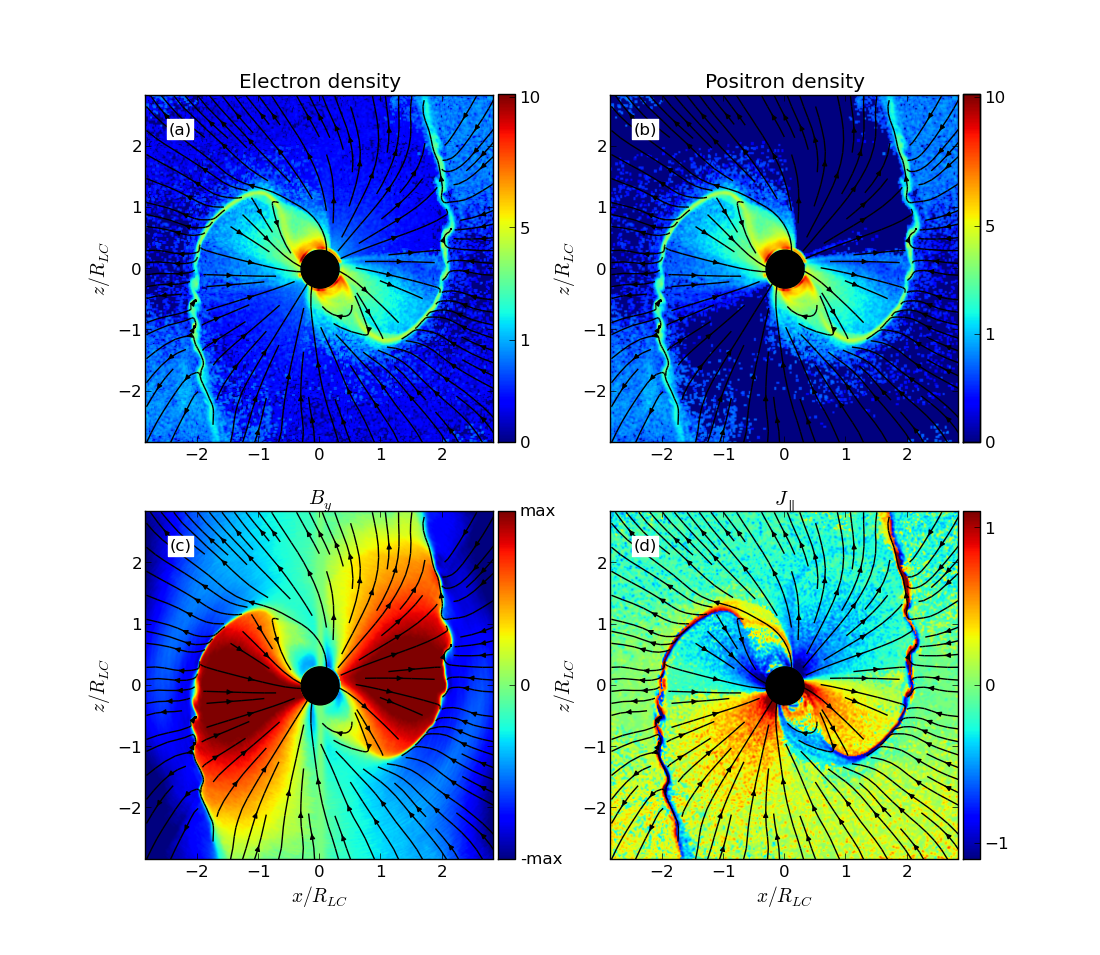}

\caption{Same as Fig.~\ref{fig1:ffree}, but for $\chi=60^{\circ}$ pulsar magnetosphere with self-consistent pair formation.}

\end{figure*}

\begin{figure*}

\centering
\label{fig4:90}
\includegraphics[width=1.\textwidth]{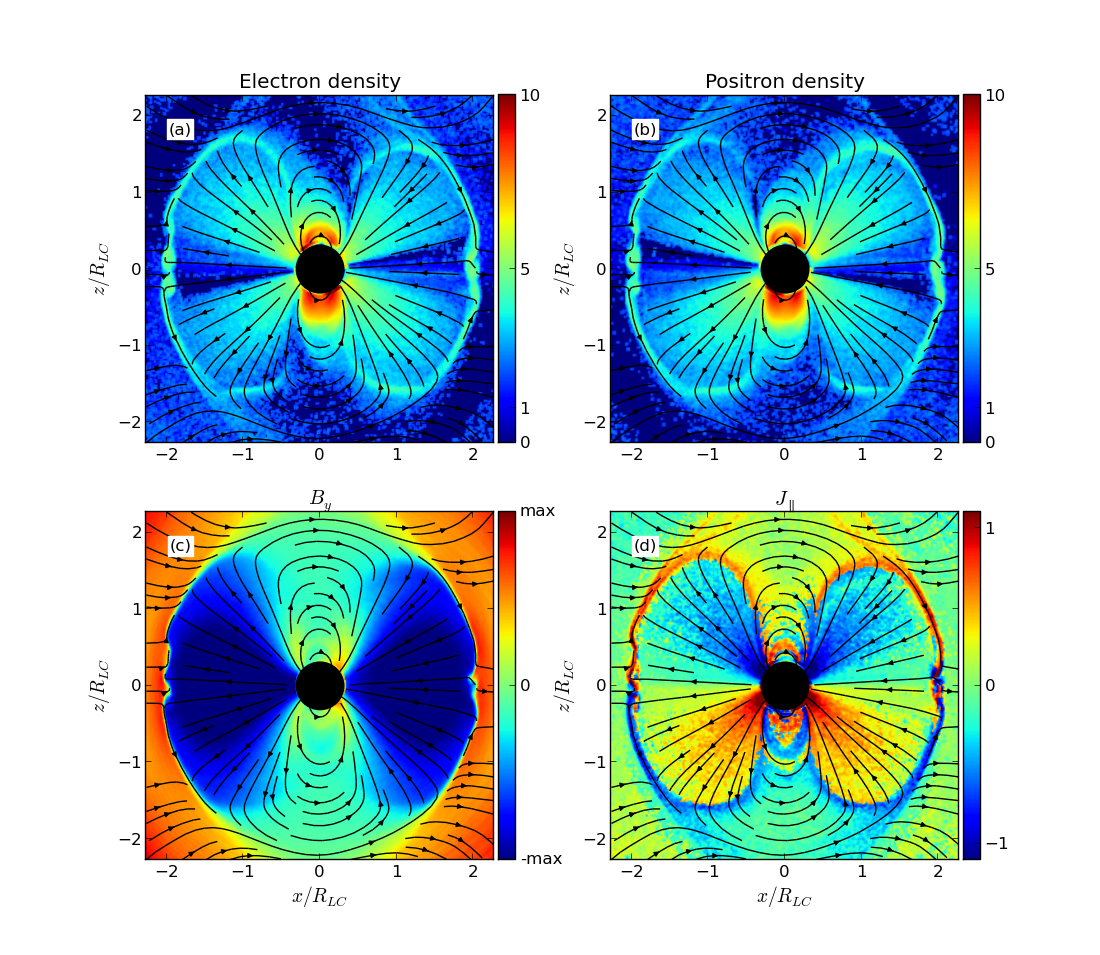}

\caption{Same as Fig.~\ref{fig1:ffree}, but for  orthogonal pulsar magnetosphere with self-consistent pair formation.}

\end{figure*}

In order to mimic the surface charge injection, we inject neutral
plasma everywhere on the star with a rate of $0.2\Sigma=0.2(E_r(r_*)-E_r^{cor}(r_*))/4\pi$ per time step, where $\Sigma$ is the local
surface charge, $r_*$ is the injection point at the stellar surface, and $E_r^{cor}$ is the
$r$-component of the corotation electric field. We found that, in general, $E_{\parallel}$ is
not effectively screened at the surface for lower
injection rates, while larger rates lead to virtual cathode oscillations. 
Even though we inject neutral plasma, the particles are injected at rest, and one
sign of charge is pulled into the star, while the other is accelerated
into the magnetosphere.

In real pulsars, the particles that are accelerated above the polar cap emit high-energy
curvature photons that produce pairs via magnetic
conversion. The pairs are created in non-zero Landau levels
and emit synchrotron radiation that might also be converted into
pairs. In addition to magnetic conversion of photons, the
electron-positron pairs may be produced in $\gamma-\gamma$ collisions. 
In this study we ignore the mean free
path of photons and produce secondary pairs at the location of 
primary energetic particles, whenever the energy of a 
particle exceeds the threshold $\gamma > \gamma_{min}$. The threshold value is set to 
$\gamma_{min}=0.02\gamma_{0}=40$, where $\gamma_{0}$ is the full vacuum
potential drop between the pole and the equator, $\gamma_0=(\Omega_* R_{*}/c)(B_0 R_{*}/c^2)\approx 2000$. The threshold does not depend on the distance from the star. The rate of pair creation per particle in
our simulation is $2\gamma/\gamma_{min}$ per time step.
 These simulation parameters help ensure sufficient pair
supply in the regions of  magnetosphere where pair production is possible. 
In reality, in addition to particle energy, the threshold and pair yield of both the magnetic conversion and $\gamma-\gamma$ collisions depend on the strength and curvature of the local magnetic field. Our aim is not to model the true multiplicity of the flow, but to identify the regions of active pair formation that can yield high multiplicity. By using a pair production prescription that does not depend on the local magnetic field, we obtain an upper limit on the number of such active regions. 
In order to resolve the skin depth and minimize computational cost, we put a limiter on the number of
produced pairs, so that  multiplicity in every cell where pairs are produced does not exceed $10$. 
The secondary pairs are injected in the direction of primary particle motion, and are randomly distributed in the computational cell of the parent.  Their Lorentz factor is set to $4$, much less than that of the parent particle. We subtract the energy of produced pairs from the energy of the parent particle, accounting for the effect of radiation reaction force.

\subsection{Aligned rotator}

\begin{figure}

\centering
\label{fig5:loss}
\includegraphics[width=0.52\textwidth]{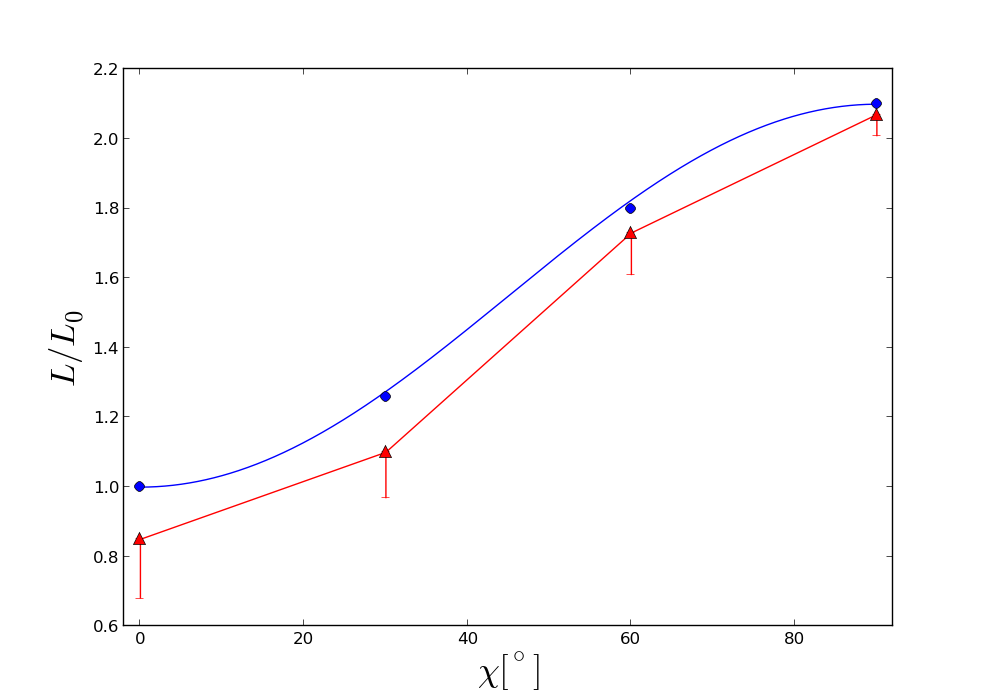}

\caption{Poyting flux luminocity of solutions with realistic pair formation in units of
  $L_0=\mu^2\Omega_*^4/c^3$ as a function of the inclination angle
  (red triangles), measured at the stellar surface. The error bars correspond to the dissipated fraction of the Poynting flux within $2R_{LC}$. Blue
  points show the results of PIC simulations with abundant pair
  formation (\S \ref{sec:test}), and the blue curve shows
  prediction of the MHD model.}
\end{figure}

The magnetosphere of the aligned rotator with self-consistent pair formation
is shown in Fig.~2c. Its magnetic field structure is close to the 
force-free solution: field becomes mostly toroidal  beyond $R_{LC}$, and poloidal field
lines stretch and become radial. The opposite polarities of 
toroidal field are supported by the equatorial current sheet. We observe moderate kinking of the sheet
beyond $2R_{LC}$, that does not affect the Y-point region during our simulation.

The polar region is inactive to pair formation,
as pointed out by \citet{Tim13} and CB14. This happens because 
the current flow is sub-GJ (see Figure 2d), thus $E_{\parallel}$ within the light cylinder is easily screened by sub-relativistic electron outflow shown in Figure 2a. 
However, pair production does take place in the return current layer
and in the current sheet, which
have high multiplicity of secondary pairs as shown in Fig.~2a,b. The acceleration of parent particles occurs due to strong $E_{\parallel}$ in the current sheet. The electric field points away from the star, and  positrons are accelerated outwards, while  high-energy electrons stream towards the star. These  electrons pass through the Y-point and penetrate into the inner magnetosphere by a few Larmor radii. The current
sheet width scales with magnetic field strength at the light
cylinder, staying of the order of average Larmor radius of particles. 
Most of the open field lines do not cross the null surface, and the
electron outflow on these lines extends to infinity. The last open field lines, however, go through 
the null surface. Since the electron outflow cannot cross the null surface, a 
vacuum gap opens above the current sheet (cf.~CB14). The spin-down of this solution, measured as the flux of
 electromagnetic energy at the stellar surface, is $\approx 0.85\mu^2\Omega_*^4/c^3$.

We note that confining pair creation to 
inner magnetosphere, $r\leq 2R_*$, or increasing the global pair
creation threshold to
values comparable with $\gamma\approx\sigma_{LC}$, where $\sigma_{LC}$
is the magnetization parameter at the light cylinder, leads to the suppression of return current, and the magnetosphere relaxes to the low
multiplicity solution (\citealp{Benoit14}; CB14). In this
state the magnetosphere consists of an equatorial disk of positrons and
a polar dome of electrons. Positrons are pulled from the tip of the disk
by strong electric field, $E>B$; however, the resulting current  is not
large enough to substantially modify the dipolar magnetic field structure.

\subsection{Oblique rotator}

We found that magnetospheres of rotators with inclination angles
$\chi \lesssim 40^{\circ}$ are qualitatively similar to the aligned
solution, showing no pair creation activity in the polar cap
outflow. The bulk outflow in these solutions is charge-separated, and pair production happens only in the current sheet and the return layer. These solutions relax to the oblique disk-dome structure if the pair
production in the outer magnetosphere is suppressed. This happens
because the return
current layer, where pair production is possible, cannot be formed by particles lifted from the
surface, since the sign of the necessary current is opposite to the sign of available
charges. As there is no current outflow, the oblique disk-dome solutions spin down at the vacuum rate.
 
Solutions with high obliquities, $\chi \gtrsim 40^{\circ}$, do show pair creation in the polar cap zone. There are
regions on the polar cap with volume return current  
exceeding the GJ value \citep{Tim13}. The structure of the $60^{\circ}$
rotator is shown in Figure 3c. The magnetospheric structure is similar
to what we found in \S \ref{sec:test}, with closed and open
zones, Y-point and the current sheet. Pairs are produced in the region
of volume return current and in the current sheet. The flow in the sheet is supported not only by the 
local pair creation, as in the case of the aligned rotator, but also by the 
reconnection-induced inflow of charges produced in the polar cap discharge. The plasma density is
not uniform in the polar cap as seen from Figures 3a and b. Part
of the polar cap supports charge-separated electron outflow with $\alpha=j_{\parallel}/j_{GJ} \approx 0.95 < 1$. The
Lorentz factor of this flow is $\approx 25$,  independent of the surface magnetic field. This is consistent with the estimate $\gamma \approx 2\alpha/(1-\alpha^2)$ for the space-charge-limited flow (\citealt{Beloborodov08,Tim13}) in the case of sub-GJ current. The volume return current region
shows active pair production. The polar discharge has
quasi-periodic behavior on the timescales short compared to the pulsar
rotation period, with episodes of efficient pair production followed by quiet states. The
three-dimensional study of the cascade intermittency is
important for modelling the pulsar radio emission and will be
discussed elsewhere. The multiplicity of the flow in the active region of the polar cap is
$\approx 10$, close to the limiting value. We performed a simulation
with twice larger limiting value and observed that the multiplicity of
the flow increases by roughly the same factor. Though not the whole polar cap shows active pair formation, on average
the pulsar wind is supplied with plasma of high multiplicity,
concentrated in the equatorial current sheet wedge.
The current sheet beyond the light cylinder should thus be the main source of $\gamma$-photons, and the polar cap
discharge is a candidate for the ``bridge'' emission. As in the
aligned case, mainly positrons are accelerated in the current sheet of the $60^{\circ}$
rotator. We find that current sheet of the pair forming solution shows more
efficient kinking than the test solution from \S \ref{sec:pulsar}, though it does not strongly affect the global magnetospheric structure. 

We find that the volume return current region in our simulations
is bounded by the null surface that crosses the polar cap; this current is carried by
positrons. However, in the limit of very small star $R_*/R_{LC}
\to 0$ the null surface is
located outside the polar cap for inclinations that are not too close to
$90^{\circ}$, and the outflowing current is supposed to be carried only
by electrons. Thus, the pair-producing region with $j>j_{GJ}$ may disappear\footnote{However, the value of  $j_{\parallel}/j_{GJ}$ in the bulk region for the small
stellar radius may also change.}.
We conclude that precise obliquity angle for transition between active and inactive state of the polar cap may change for $R_*\ll R_{LC}$. Although our present
simulations are already physically relevant for millisecond pulsars with high
$R_*/R_{LC}$, we will investigate the smaller values of $R_*/R_{LC}$ in future work. 

The $90^{\circ}$ rotator shows efficient pair production in the
whole polar cap (Fig.~4). In contrast to the aligned rotator, the
current flow on the polar cap is anti-symmetric with respect to the
equatorial plane. Current density exceeds $j_{GJ}$,
necessitating the presence of $e^{\pm}$ discharge.  Pair formation also happens in
the current sheet. Moderate kinking of the sheet is observed close to the equatorial plane.

In Figure 5 we present the spin-down of our oblique solutions with
pair formation, measured at the stellar surface. In our solution for the aligned rotator with pair discharges  
approximately 20\% of the Poynting flux is dissipated within
$2R_{LC}$. The dissipated fraction of total energy losses decreases for larger inclination angles and reaches 3\% for the orthogonal rotator. The
dissipated energy is converted into particle energy in acceleration regions. The deviation of spin-down
from solutions with abundant pair formation discussed in \S3 is small and becomes
negligible with increasing inclination. The origin of this deviation is due to the presence of vacuum-like
regions inside the light cylinder that are not in corotation with the star.

\section{Discussion}

In this Letter we presented the magnetospheric structure of oblique pulsars with pair formation.
We found that pulsars with low obliquities produce active
solutions only if pair formation in the outer magnetosphere is working.
In these solutions the pairs are produced in the equatorial current sheet beyond the light cylinder and in the return current layer. If the
radio emission is produced in the high multiplicity flow at the polar
cap, these pulsars should have mainly double-peaked profiles
associated with return current layers.
Pulsars with obliquity angles $\gtrsim 40^{\circ}$ show significant pair production activity
in the polar regions. Their radio beams may have multi-peaked structure. The spin-down
of pair solutions is close to MHD model predictions.

Our prescription for pair formation is highly idealized. For simplicity we neglected the mean free path of photons and assumed the pair formation threshold to be independent of the local conditions. While this approximation is reliable for the polar cap discharge, the $\gamma-\gamma$ process which can account for the activity in the current sheet is intrinsically non-local, which requires more careful
treatment. Two processes may contribute to launching
the $e^{\pm}$ discharge in the current sheet: collisions of
energetic synchrotron photons emitted by counter-streaming particle
flows, and collisions of thermal X-ray photons from the neutron star surface, which may be up-scattered by energetic particles in the sheet, with synchrotron photons. Both mechanisms should operate most efficiently close to the
Y-point, where magnetic field in the sheet is strongest, and the emitted
synchrotron photons are the most energetic. We assumed efficient pair formation in the current sheet, which may not happen in pulsars with low magnetic field strength at the light cylinder. In low obliquity pulsars with no pair formation in the current sheet, the return current layer is suppressed, and such pulsars should not shine in the radio band, unless some physics beyond our model can reestablish pair formation. 

A potentially important complication may arise due to the restricted ability of ions to produce pairs. 
Though ions can still emit curvature photons capable of producing pairs in $\gamma-\gamma$ collisions with stellar photons, the resulting multiplicity of the flow due to this process is uncertain. This may affect the multiplicity of the polar outflow in the orthogonal rotator, where ions are extracted in the half of the polar cap.

The poloidal field at the stellar surface may be more complicated than
the pure dipole field considered in this work due to the operation of Hall effect in the
crust (e.g., \citealp{Hall}). The resulting multipolar field can modify the operation of the
polar discharge, producing more regions with efficient pair
production. 

The value of $\alpha=j_{\parallel}/j_{GJ}$ that is essential for operation of polar discharge
may be sensitive to general relativistic corrections. In particular, the value of GJ charge density is reduced due to Lense–-Thirring effect (\citealt{VSB90, MTsyg92}). The increased value of $\alpha$ may lead to the ignition of polar discharge in low obliquity pulsars. This will be investigated in future work. 

We thank Jonathan Arons, Vasily Beskin, Alexander Tchekhovskoy, and Andrey Timokhin for fruitful discussions. This research was supported by NASA grants NNX12AD01G and NNX13AO80G, Simons Foundation (grant 267233 to AS), and was facilitated by Max Planck/Princeton Center for Plasma Physics. The simulations presented in this article used computational resources
supported by the PICSciE-OIT High Performance Computing Center and Visualization Laboratory.

\end{document}